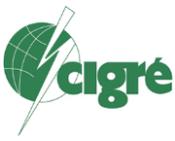

# Best Practices for Large Load Interconnections: A North American Perspective on Data Centers

Rafi Zahedi, Amin Zamani, Rahul Anilkumar
Quanta Technology, LLC

## SUMMARY

Large loads are expanding rapidly across North America, led by data centers, cryptocurrency mining, hydrogen production facilities, and heavy-duty charging stations. Each class presents distinct electrical characteristics, but data centers are drawing particular attention as AI deployment drives unprecedented capacity growth. Their scale, duty cycles, and converter-dominated interfaces introduce new challenges for transmission interconnections, especially regarding disturbance behavior, steady-state performance, and operational visibility.

This paper reviews best practices for large-load interconnections across North America, synthesizing utility and system operator guidelines into a coherent set of technical requirements. The approach combines handbook and manual analysis with cross-utility comparisons and an outlook on European directions. The review highlights requirements on power quality, telemetry, commissioning tests, and protection coordination, while noting gaps in ride-through specifications, load-variation management, and post-disturbance recovery targets. Building on these findings, the paper proposes practical guidance for developers and utilities.

## KEYWORDS

Artificial intelligence, data center, grid interconnection, large loads, interconnection requirements

## 1. INTRODUCTION

With the rapid transformation of the North American electric grid, emerging large loads, such as data centers, cryptocurrency mining, hydrogen production, and heavy-duty EV charging, are requesting interconnections at unprecedented scale and speed, often exceeding today's largest operating facilities. Early events associated with these types of loads already highlight reliability risks: in July 2024, a 230 kV fault in the Eastern Interconnection tripped 1,500 MW of large data center loads, briefly raising frequency and voltage [1]–[2], [3]. Between late 2023 and early 2025, ERCOT and the Eastern Interconnection recorded 25 crypto-mining load losses of 100–400 MW each [2].

Artificial intelligence (AI), cloud services, and high-performance computing are among the primary drivers of rapid data center expansion. In 2024, global data generation exceeded 149 zettabytes, fueled by AI-enabled applications, widespread digitalization, and the proliferation of connected devices [4]. The U.S. leads with 40% of global capacity, as electricity demand from data centers rose from 58 TWh in 2014 to 176 TWh in 2023, with the US Department of Energy projecting 325–580 TWh by 2028 [5]. During this period, workloads shift toward inference (80% to 85%) and diversify geographically, with edge computing projected to handle half of AI processing by 2028 [7].

Data centers are a distinct class of large electrical load. Unlike crypto mining or traditional industrial processes, their demand is computationally driven with a constant baseline but sudden spikes from AI training and real-time inference [6]. These bursty patterns make data centers uniquely challenging for maintaining power quality and ensuring grid stability and integration.

Interconnection requests are now outpacing legacy processes, creating reliability, study, and process challenges for utilities and Independent System Operators (ISOs). From a performance standpoint, fast ramps and step changes can trigger large-scale disconnections if ride-through and reclosing requirements are not properly defined [7]. Without adequate filtering, power-electronic interfaces can introduce harmonic distortion, phase imbalance, and broader power quality concerns. These pressures underscore the need for harmonized technical requirements at the point of interconnection (POI).

This paper provides a comparative analysis of large-load interconnection requirements across U.S. power entities, with emphasis on technical requirements. By extracting explicit numeric limits from utility and ISO guidance, the study highlights areas of convergence and divergence in current practice. The work contributes a framework and visual benchmarks designed to support both policy development and practical planning for integrating data center loads.

## 2. USA LARGE LOAD TECHNICAL INTERCONNECTION REQUIREMENTS

The operational concerns of large loads are distinct from traditional motor-dominated load. To ground the discussion in practice, this section collates utility and ISO requirements into discipline-specific primers, followed by comparative tables that report the most prescriptive, numeric obligations where they exist.

Figure 1 groups the interconnection requirements into eight broad categories: load management, power quality, voltage ride-through (VRT), frequency ride-through (FRT), post-disturbance recovery, protection requirements, operational communications and control with compliance, and study and modeling requirements. Table 1 compiles guidance from AEP, Dominion, ERCOT, Ameren (MISO), PG&E, Southern Company, and SPP, using a POI-based frame that emphasizes explicit numerical limits and restoration targets. This structure allows direct comparison while retaining each entity's terminology and scope.

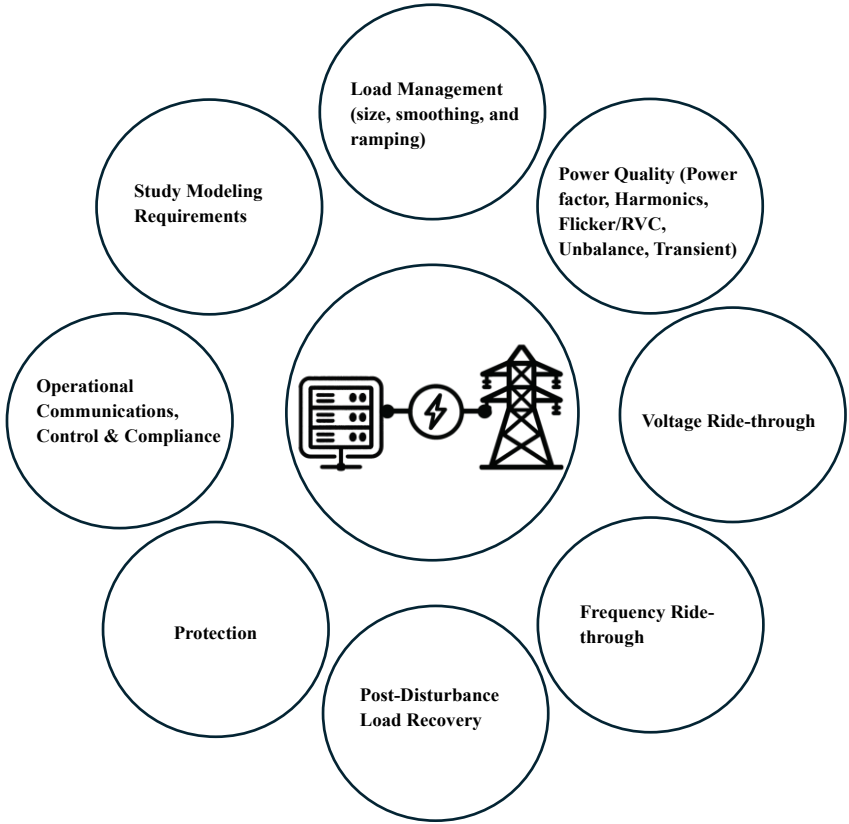

**Figure 1. Technical Requirements Categories Investigated in This Paper**

## 2.1. Load Management (size, smoothing, and ramping)

As noted earlier, AI training facilities create rapid swings in demand, underscoring the need for clear POI envelopes and staged restoration profiles. Interconnection requirements should therefore link step and ramp limits with reconnection staging, supported by high-resolution telemetry and validation to align operational control with the documented risks of large loads [1]. Further, smoothing strategies are required to dampen fluctuations and stabilize load curves [8].

Entities differ considerably in how they define "large load," as shown in Figure 2. Dominion sets a threshold for power-electronic interface large loads at ≥50 MW, while ERCOT classifies ≥75 MW at a single site as large and applies a 25 MW visibility threshold at substations. This means that if multiple facilities behind the same transmission substation together exceed 25 MW, ERCOT requires enhanced observability (e.g., telemetry, status, coordination) even if each facility is below 75 MW. Southern Company defines transmission-connected (>40 kV) large loads as ≥50 MW, and SPP designates high-impact large loads as >10 MW at 69 kV or >50 MW at higher voltages. By contrast, AEP, Ameren (MISO), and PG&E do not set a fixed MW threshold.

As shown in Table 1, Southern Company is the only company that enforces a numeric cap of ≤ 20 MW/min for load ramping under normal operation. AEP requires submission of a Load Ramp Schedule, ERCOT mandates a Load Commissioning Plan consistent with Large Load Interconnection Study limits, Dominion specifies a site-specific staged reconnection plan, and SPP calls for constant-current (rather than constant-power) control during disturbances. Other entities focus more on planning data, control-center coordination, and forecasting rather than prescribing explicit caps.

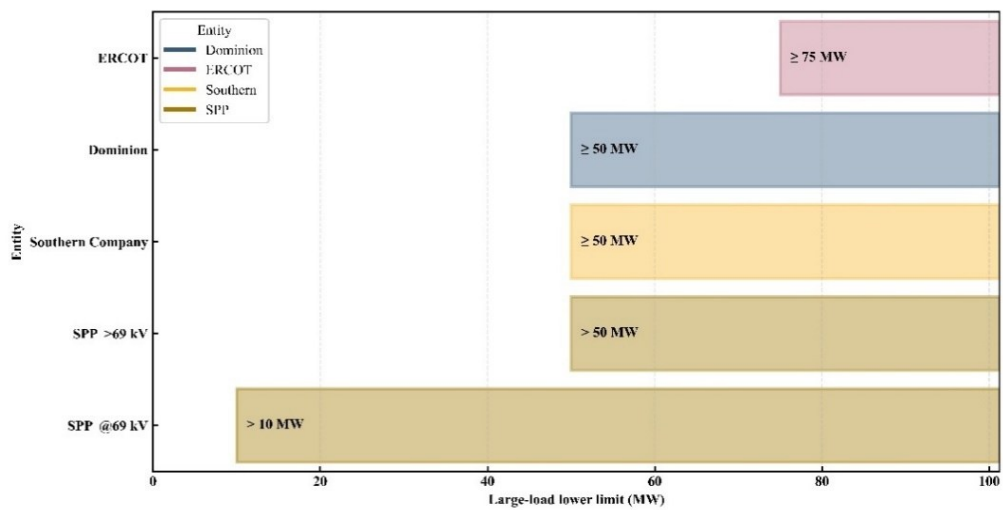

Figure 2. Large Load Definition by Different Entities

## 2.2. Power quality

Interconnection requirements generally pair explicit power quality (PQ) limits at the POI with monitoring provisions so disturbances can be detected before affecting other customers or system performance [9]. Table 1 shows that Dominion requires ≤3% instantaneous voltage fluctuation at the POI along with permanent metering that records time-stamped, high-resolution logs for compliance audits. Ameren adopts IEEE 519 for harmonics mitigation and IEEE 1453 for flicker and rapid voltage change (RVC). PG&E enforces Rule 2 limits (THD ≤5%, ~10% imbalance), CAISO applies power factor limits at POI, and Southern Company requires harmonic spectrum out to the 50th order, permanent monitors, and enforceable mitigation. AEP references a designated measurement point with IEEE 1453 for flicker/RVC.

Among these, only Dominion and Southern Company explicitly require permanent PQ monitoring at the POI. Ameren may install monitors but does not mandate permanent recorders, PG&E enforces PQ limits without specifying devices, and AEP identifies a measurement point without prescribing a permanent meter. ERCOT and SPP, based on the publicly available materials, do not require load-side PQ monitoring.

Table 1. Summary of comparative analysis of load management and power quality for large load interconnection

| Entity | Category | |
|---|---|---|
| | **Load Management (size, smoothing, and ramping)** | **Power Quality** |
| **AEP** [10] | • Definition: AEP treats projects on a case-by-case basis.<br>• Customer submits a Load Ramp Schedule describing step magnitudes, cadence, reconnect/dropout thresholds, and delays; AEP checks steps/ramps against flicker and RVC planning limits and may require mitigation. | • PQ is assessed at an AEP-designated point of measurement (often the POI).<br>• Flicker/RVC must meet IEEE 1453; harmonic expectations are reviewed for modern electronic equipment. |
| **Dominion** [11] | • Definition: ≥50 MW but may designate <50 MW based on characteristics or location.<br>• Customer should provide site-specific data (load type and transfers) and a staged reconnection plan. | • Instantaneous voltage fluctuation at the POI ≤ 3%.<br>• Permanent POI PQ metering with time-stamped logs (P, Q, PF, harmonics to 50th, Pst/Plt, RVC, unbalance, transients, frequency) and ≥90-day retention. |
| **ERCOT** [12], [13], [14], [15] | • Definition: ≥75 MW at a single site; visibility threshold ≥25 MW at a common substation.<br>• Behavior governed by the commissioning plan, set by the transmission service provider based on study limits; no explicit step/ramp is defined. | NA |
| **MISO-Ameren** [16] | • Definition: not a fixed MW threshold; procedures apply to any transmission-connected end user.<br>• Application includes MW/MVAr/PF, max demand, actual peak (if applicable), 10-year forecast, and unique operating characteristics. | • Harmonics must meet IEEE 519; Ameren may install POI meter; customer must mitigate if limits are violated.<br>• Flicker/RVC assessed to IEEE 1453; e.g., 2.0% critical-bus dip is marginal and may require mitigation.<br>• PF outside 95% lag–95% lead requires customer-side correction. Persistent PQ non-compliance can lead to disconnection. |
| **PG&E** [17], [18], [19], [20], [21] | • Definition: not specified; no step/ramp caps are published.<br>• All connections/separations are coordinated with the PG&E control center; remedial action schemes may be required if studies identify them. | • Enforces Electric Rule 2 at the POI; total voltage THD ≤ 5% and the customer funds mitigation.<br>• Phase balance: difference between any two phases at peak load should not exceed ~10%.<br>• CAISO PF band is 97% lag–99% lead. |
| **Southern Company** [22], [23] | • Definition: ≥50 MW at >40 kV; SCT (Southern Company Transmission) may designate smaller facilities based on characteristics or location.<br>• Maximum active-power ramp ≤ 20 MW/min under normal operation; settings are reviewed and demonstrated; energization inrush and steps must respect RVC limits. | • Comply with SCT's PQ policy; stricter limits may be imposed from study results.<br>• Provide expected harmonic spectrum to the 50th order; SCT may require mitigation and can temporarily reduce or disconnect the load.<br>• SCT installs permanent POI PQ monitoring; PF must remain within coordinated limits. |
| **SPP** [24] | • Definition: >10 MW at 69 kV or >50 MW at higher voltages.<br>• For electronic loads, operate in constant-current mode during disturbances and avoid constant-power mode; no steady-state step/ramp defined. | NA |

### 2.3. Voltage ride-through

VRT is the capability and obligation of a facility to remain connected at the POI through specified combinations of voltage magnitude and duration that occur during faults, using POI-referenced envelopes and verified by standardized test methods [25]. For data centers, front-end rectifiers and limited dc-link energy make constant-power behavior prone to tripping on deep, short sags unless buffered by an Uninterruptible Power Supply (UPS) systems or coordinated controls; this is why power entities are moving to explicit load VRT expectations [26].

As illustrated in Table 2, SPP has published a time-voltage curve (0.90–1.10 pu continuous, with short-duration tolerances down to ~0.50 pu for ~0.15 s) and transient overvoltage withstand aligned with IEEE 2800. Dominion emphasizes reclosing tolerance instead, requiring ride-through of several reclosing shots (each ~50–70 ms) with undervoltage pickup near 85%. ERCOT's proposed large-load VRT requires continuous operation between 0.90–1.10 pu, with staged ride-through outside that range based on the ITIC curve, global experience, and existing IBR requirements (see Section 3).

## 2.4. Frequency ride-through

FRT requires facilities to remain connected through defined combinations of frequency deviation and duration, typically expressed as frequency–time curves with "must-run" and "must-trip" regions [27]. For converter-dominated data centers, simultaneous tripping or transfer to UPS can remove hundreds of megawatts at once, motivating explicit FRT requirements beyond system-wide under-frequency load shedding (UFLS) schemes [26]. Most generator codes already require continuous operation near 59.4–60.6 Hz, while the emerging practice is to apply analogous POI-referenced windows coordinated with UFLS. In data center applications, FRT should also include staged recovery and high-resolution telemetry to verify compliance [27].

As shown in Table 2, SPP publishes a detailed profile (continuous 58.8–61.2 Hz, extended to 299 s, permissive trip beyond 61.8/<57.0 Hz). By contrast, most other entities embed load behavior within UFLS frameworks: AEP aligns with regional UFLS, Dominion requires participation in UFLS with necessary relays, Ameren defines staged UFLS blocks (59.3/59.0/58.7 Hz), PG&E applies UF relaying as directed by CAISO, and Southern requires verification of customer settings without publishing numeric bands.

Table 2. Summary of comparative analysis of VRT and FRT requirements for large load interconnection

| Entity | Category | |
|---|---|---|
| | VRT | FRT |
| **AEP** [10] | • No POI time–voltage envelope for loads is published; voltage performance is managed via PQ limits and settings review during studies. | • Coordination with regional UFLS (under-frequency load shedding); no separate numeric frequency-time envelope is prescribed. |
| **Dominion** [11], [28] | • Remain connected through several automatic reclosing shots (50-70 ms); recommend undervoltage pickup ≤85% with short timers to avoid nuisance transfers. | • Participation in system-wide UFLS with appropriate relays and communications. |
| **ERCOT** [12], [13], [14], [15] | • ERCOT follows ITIC Curve and IEEE 1668 for undervoltage, and NOGRR245 for overvoltage.<br>• Proposal applies prospectively. | • Numeric load frequency settings are "to be developed"; any adopted settings are to be set to maximum feasible capability. |
| **MISO-Ameren** [16] | • No load ride-through curve; undervoltage load shedding is not standard and may be applied on a case-by-case basis as an interim measure. | • UFLS stages near 59.3, 59.0, and 58.7 Hz (~10% each).<br>• New connections may initially be allowed without UFLS, but Ameren can require UFLS at a later retrofit. |
| **PG&E** [17], [18], [19], [20], [21] | • No load time-voltage curve; unexpected separations are reported with causes and event files; settings should be coordinated with PG&E. | • No explicit load frequency window; under-frequency relaying is applied where applicable, with telemetry; PG&E interfaces under ISO direction in emergencies. |
| **Southern Company** [22], [23] | • Submit voltage ride-through settings for SCT review and adjust as directed or per standards; settings to be verified through testing/monitoring. | • Submit frequency ride-through settings for SCT review and adjust as directed or per standards; settings to be verified through testing/monitoring. |
| **SPP** [24] | • Follows ITIC curve and IEEE 1668-2017, IEEE standard 2800-2022 (Clause 7 adopted by SPP), and NERC standard PRC-029-1. | • Example withstand: continuous 58.8–61.2 Hz; extended tolerance for 299 s outside that band; trip permissible above ~61.8 Hz or below ~57.0 Hz. |

## 2.5. Post-disturbance load recovery

In this paper, post-disturbance load recovery refers to the requirements for reconnecting data center load to the grid following a fault or power quality event. This requirement is critical to preserve grid integrity, since large portions of data center load may temporarily shift to backup power during a disturbance. Table 3 shows that ERCOT is the only entity specifying a numeric target: ≥90% of pre-disturbance load restored within 1 second once $V_{POI} \geq 0.9 pu$. AEP and Dominion require customer-specific restoration plans (thresholds, delays, block sizes, ramps), but no uniform limits. Ameren coordinates recovery through switching and reactive support. PG&E requires control-center approval before reconnection, including safeguards against unattended remote restore. Southern Company enforces approved procedures and validated ramp limits, with curtailment capability, while SPP expects recovery consistent with its VRT/FRT boundaries and operating guidance.

## 2.6. System protection

Table 3 summarizes practices: AEP requires complete protection data and clearing verification; Dominion specifies delta high-side intertie transformers and may reduce service if harmful interactions occur; ERCOT requires remotely operable interrupters and includes system oscillation analysis in interconnection study; Ameren specifies redundant schemes and breaker-failure coverage; PG&E mandates communication-aided schemes, periodic testing, and prohibits ≥100 kV taps; while Southern Company adds instrument transformer accuracy, breaker ratings, and potential inclusion in load shedding schemes. SPP does not provide detailed load-side protection rules. Nearly all entities mandate utility review and approval of the protection system before energization.

Table 3. Summary of post-disturbance load recovery and protection requirements for large load interconnection

| Entity | Category | |
|---|---|---|
| | **Post-Disturbance Load Recovery** | **Protection** |
| **AEP** [10] | • Customer provides staged reconnection and ramp plans (thresholds, delays, block sizes, ramps); AEP coordinates to avoid secondary dips and RVCs. | • Customer supplies full protection data; AEP confirms interrupting duties, clearing times, and transformer/high-side/low-side device adequacy. |
| **Dominion** [11], [28] | • Customer specifies reconnection method and ramp process; numeric recovery level/time is not prescribed. | • Intertie transformer high-side is normally delta; configurations require review.<br>• Reduction/disconnection may be required if interactions are harmful; inter-substation transfers need approval. |
| **ERCOT** [12], [13], [14], [15] | • Proposed target (Area B): recover to ≥90% of pre-disturbance consumption within 1 second once $V_{POI}$ ≥ 0.9pu; if disconnected (Area C), restoration is staggered across multiple large loads. | • Each POI includes remotely operable disconnect devices able to interrupt fault current and isolate the load; the study scope includes steady-state, stability, short-circuit, and Subsynchronous Oscillation (SSO) considerations. |
| **MISO-Ameren** [16] | • No numeric minimum level/maximum time; restoration should be coordinated with Ameren using switching, reactive adjustments, and operator direction. | • At minimum, a ring-bus is required; large hubs may use a straight bus or breaker-and-a-half configuration.<br>• Breaker failure protection for customer interrupting devices; visible isolation at the POI; redundant schemes that detect faults on both sides. |
| **PG&E** [17], [18], [19], [20], [21] | • After any separation, the operator notifies the Control Center and obtains permission before reconnecting; unattended remote restoration requires verification that the circuit is energized from a PG&E-approved source. | • Utility-grade, PG&E-approved relays coordinated with PG&E line protection; visible load-interrupting device at the POI; fault-interrupting devices located close to POI.<br>• Communication-aided line schemes and transfer-trip channels may be required; relay/breaker testing before energization and every six years.<br>• New taps on ≥100 kV lines are not permitted; interconnect at a substation. |
| **Southern Company** [22], [23] | • No utility-wide numeric minimum level/max time; restoration follows approved operating procedures and validated ramp limits; curtailment capability and ramp-rate performance are checked before "Full Readiness." | • Utility-grade relays with coordinated settings; relaying-accuracy CTs; SCT determines intertie breaker rating and may require high-speed protection communications.<br>• High-side transformer normally delta; on-site generation is interlocked to prevent parallel operation unless allowed.<br>• Load may be included in load shedding schemes. |
| **SPP** [24] | • No numeric minimum level/max time; expectation is to remain connected within V/F boundaries and coordinate restoration per operational guidance. | NA |

## 2.7. Control and communication requirements

For large loads, utilities are moving toward substation-grade communications and control to ensure fast, deterministic signaling for operations and auditable data for compliance. As summarized in Table 4, visibility expectations are rising: Dominion mandates permanent POI PQ meters with ≥90-day logs; ERCOT requires ≤10-s telemetry, continuous state reporting, and redundant ICCP; Ameren requires continuous MW/MVAr/V/I telemetry over Harris 5000 or equivalent; PG&E specifies 24/7 communications with operator notification, cellular/Ethernet revenue meters, and dedicated protection comms as needed; Southern Company requires company-owned RTUs with validated points and may deploy Phasor Measurement Units (PMU) and/or advanced Digital Fault Recorders (DFR); AEP calls for reliable telemetry but does not specify a protocol.

## 2.8. Study modeling requirements

Data center loads behave as constant-power loads, drawing more current when the voltage decreases and less when the voltage increases. This behavior produces negative incremental impedance, which can amplify disturbances and compromise stability during large transients. Therefore, planners have begun supplementing positive-sequence studies with Electromagnetic Transient (EMT) analyses, particularly when local short-circuit strength is low [29].

Table 4 shows that all entities require steady-state and short-circuit studies, with EMT analysis added as needed. AEP accepts PSS/E and may request PSCAD/EMTDC models; Dominion requires PSS/E with as-planned/as-built data and may also request PSCAD models; ERCOT defines scope and model registration through interconnection studies; Ameren adds short-circuit, stability, switching, or EMT studies under a study agreement; PG&E cites ASPEN, PSLF, PSCAD, and RTDS (at 500 kV) studies; Southern Company requires as-planned/as-built/as-left datasets with simulation-based validation; and SPP emphasizes dynamic load-model validation through modeling, commissioning tests, and operational monitoring.

**Table 4. Summary of communications, control, compliance, and study requirements for large load interconnection**

| Entity | Category | |
|---|---|---|
| | **Operational Communications, Control & Compliance** | **Study Modeling Requirements** |
| **AEP** [10] | • Reliable operational telemetry is required for real-time coordination and compliance; no specific communication protocol is mandated. | • Steady-state, short-circuit, and (as needed) stability studies; inputs include step-change cadence, harmonic spectra, VAR devices, and multi-year forecasts.<br>• Positive-sequence models in PSS/E are accepted; EMT studies in PSCAD may be requested where needed. |
| **Dominion** [11], [28] | • Permanent POI PQ meters; time-aligned, high-resolution logging with ≥ 90-day retention.<br>• Facilities subject to UFLS must maintain communications to coordinate actions. | • Provide CMLD/EV or user-defined models in PSS/E with as-planned/as-built data; PSCAD may be requested for interaction/stability studies. |
| **ERCOT** [12], [13], [14], [15] | • Operational telemetry to ERCOT at ≤10-s scan rate with condition detection; continuous breaker/switch state is needed.<br>• Fully redundant ICCP links with automatic failover. | • A large load must pass steady-state, stability, and short-circuit studies (plus any other studies the TSP/ISO deems necessary).<br>• Additional load at the same site is not included in ERCOT's planning/operational models until the current request's studies are finished, and agreement executed.<br>• No specific software is mandated in the materials provided. |
| **MISO-Ameren** [16] | • Continuous telemetry (MW, MVAr, current, voltage) using Harris 5000 protocol or agreed-upon equivalent; 24-hour contacts; annual meter testing; customer maintains telecom circuits. | • Ameren begins with power flow studies; it adds short-circuit, stability, switching, or EMT as needed under a study agreement, consistent with NERC TPL-001; no specific commercial tool is mandated. |
| **PG&E** [17], [18], [19], [20], [21] | • 24-hour comms with Control Center; telephone service required; revenue meters use cellular modem and, when possible, Ethernet with static IP.<br>• Dedicated comms for protection, such as transfer trip as specified by PG&E. | • Modeling submittal is required; PG&E cites ASPEN, PSLF, PSCAD models (and RTDS model for 500 kV).<br>• Pre-energization testing and PG&E witnessing are mandatory with defined tolerances and report lead times. |
| **Southern Company** [22], [23] | • Data link from SCT-owned interconnection RTU, typically serial over fiber; real-time telemetry points exchanged and tested before energization; SCT may deploy PQ meters, PMU, and advanced DFR devices. | • Provide site-specific data at "as-planned/as-built/as-left" stages; SCT verifies via simulations before initial energization and through a performance-validation period.<br>• The appendix lists modeling inputs (load composition, PF, sag/swell and reconnection thresholds, frequency trips, ramp rates, harmonic spectrum, load profiles).<br>• No specific commercial tool is mandated; EMT analysis may be requested. |
| **SPP** [24] | • Compliance via modeling, commissioning tests, and operational monitoring; the document does not prescribe a specific communications protocol. | • Studies, commissioning tests, and monitoring emphasize dynamic load-model validation and ride-through demonstration; no specific commercial tool is mandated. |

## 3. EUROPEAN EXPERIENCE AND VISION

Across Europe, the binding framework for large-load connections is the Network Code on Demand Connection, which harmonizes rules for connecting demand facilities and distribution systems [30]. It also makes VRT expectations explicit at the POI. For Continental Europe, this means a continuous operating window of 0.90–1.05 pu and a sustained band of 1.05–1.10 pu for 20–60 minutes, as set by the transmission operator.

Figure 3 compares low-voltage ride-through envelopes applied by RTE (France), Energinet (Denmark), and EirGrid (Ireland) [31] with numeric VRT curves from two U.S. entities, SPP and ERCOT. At the system level, European studies show that congestion at legacy hubs can delay new connections by up to 13 years. To reduce queues, recommended measures include strategic siting, phased or non-firm connections, and smarter connection agreements, which could cut timelines to around a year. The analysis also highlights the need for pilots on data center flexibility and greater transparency, such as capacity maps and visible queue data, to better coordinate siting and investment [32].

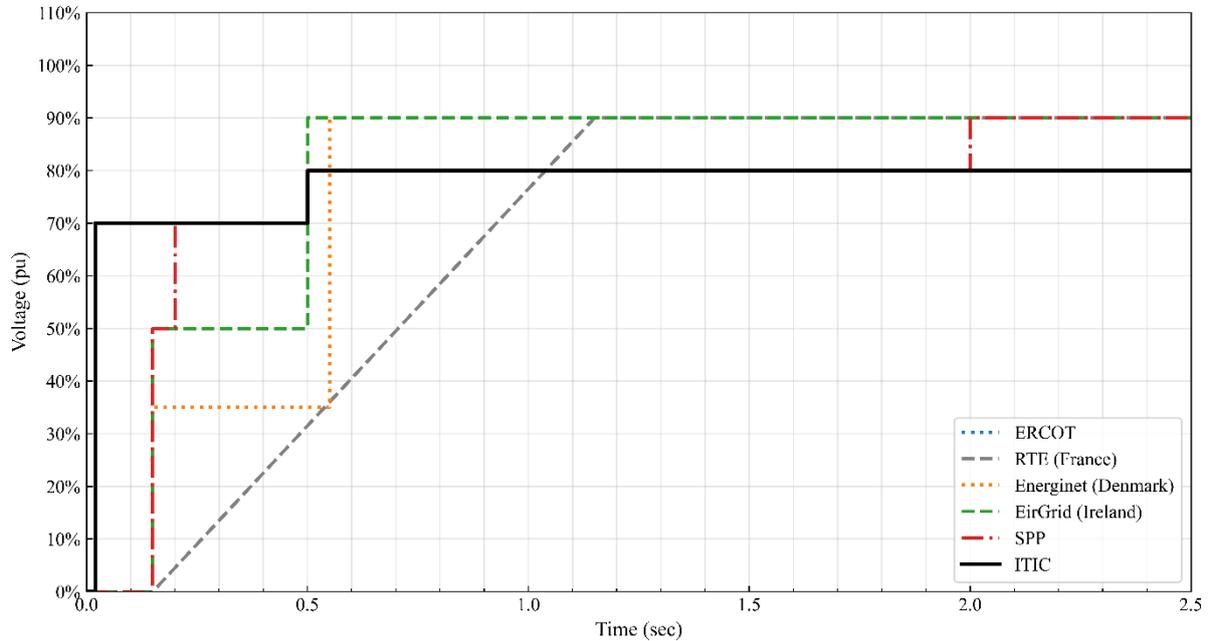

**Figure 3. VRT Curves for American and European Entities** [13], [24]**.**

### 4. CONCLUSION

This review compared interconnection requirements for large loads across U.S. utilities and ISOs, with a focus on data centers. Eight technical categories were examined: load management, power quality, voltage ride-through, frequency ride-through, post-disturbance recovery, protection, control and communications, and modeling requirements. The analysis shows uneven development across these areas. Protection requirements are relatively mature, with most entities converging on utility-grade relays, selective clearing, modeling workflows, and auditable POI isolation. Power quality expectations are also broadly aligned, referencing IEEE 519 and IEEE 1453, with growing emphasis on permanent PQ metering.

In contrast, ride-through, load recovery, and load modeling are all in progress. Only SPP and ERCOT publish explicit POI-based VRT envelopes, while most others rely on qualitative guidance. Load FRT is still tied primarily to UFLS, with limited use of numeric remain-connected windows. Post-disturbance recovery is mainly procedural, with ERCOT proposing a quantitative target. Likewise, definitions of "large load" and ramping limits vary widely across entities.

Overall, the trend points toward greater standardization of ride-through profiles, recovery targets, and visibility requirements similar to European practice. Publishing harmonized VRT/FRT envelopes, class-specific recovery benchmarks, standard telemetry expectations, and defined modeling requirements would accelerate convergence and reduce study burden. The harmonized framework and comparative analysis (Table 1–Table 4 ) provide a practical reference for utilities, regulators, and developers working to integrate large, fast-growing loads such as data centers reliably and efficiently.